\documentclass[12pt,a4paper]{article}
\usepackage{graphics,graphicx}
\usepackage{amssymb,epsfig,amsmath,euscript,array}
\usepackage{cite}

\makeatletter
\@addtoreset{equation}{section}
\makeatother

\makeatletter
\@addtoreset{equation}{section}
\makeatother

\newcounter{multieqs}

\newcommand{\be}{\begin{equation}}
\newcommand{\ee}{\end{equation}}

\begin{document}
\begin{flushright}
QMUL-PH-05-1\\
\end{flushright}

\vspace{20pt}

\begin{center}

\Large\textbf{D-brane dynamics near}
\vspace{0.2cm}
\Large\textbf{compactified NS5-branes}
\vspace{33pt}


\centerline{\normalsize \bf Steven Thomas\footnote{e-mail:
 s.thomas@qmul.ac.uk}, \
John Ward\footnote{e-mail:
 j.ward@qmul.ac.uk} }

{\em \normalsize Department of Physics\\
Queen Mary, University of
London\\ Mile End Road\\
 London, E1 4NS U.K.}

\vspace{40pt} {\normalsize\bf Abstract}

\end{center}
We examine the dynamics of a $Dp$-brane in the background
of $k$ coincident, parallel $NS$5-branes which have had one of their
common transverse directions compactified. We find that for small energy, bound orbits can
exist at sufficiently large distances where there will be no stringy effects.
The orbits are dependent upon the energy density, angular momentum and
electric field.
The analysis breaks down at radial distances comparable with the compactification radius and we must resort to using a modified
form of the  harmonic function in this region.
\newpage
\section{Introduction.}
There has been much recent study of time dependence in string theory,
particularly involving supergravity backgrounds \cite{kutasov}, \cite{burgess},
\cite{kutasov2}, \cite{sahakyan}, \cite{Ghodsi}, \cite{nakayama1}. The basic idea has been
to probe a given background geometry through the introduction of a single
$Dp$-brane, whose dynamics can be studied using an effective Born Infeld
action. Some of this work has involved probing $D$-brane backgrounds \cite{Panigrahi}, \cite{kluson3}, \cite{burgess}, \cite{burgess2}, whilst other work has focused on the
$NS$5-brane backgrounds (see \cite{kutasov}, \cite{ kluson}, \cite{kluson2}, \cite{thomas}, \cite{nakayama2}, \cite{chen})

The $NS$5-brane case is of interest since this particular
configuration breaks all the supersymmetry of type II string theory and
the BPS probe brane will be unstable. This instability can be interpreted
in terms of a rolling tachyon \cite{kutasov}, \cite{sen}, \cite{sen2}, and
so study of this system provides useful information about tachyon
condensation. In addition, since the $NS$5-brane is much heavier than the
$Dp$-brane, the source branes will curve the space around them creating an
infinite throat. As one travels down the throat, one finds that the string
coupling increases, thus we find that we are moving from the perturbative to
the non-perturbative realm.

Since we are using perturbative techniques, we
expect that our solution will break down as we approach the bottom of the
throat. However \cite{kutasov} found that there were certain values of
energy for which the solution held for some time.
Another point to consider is the energy emission associated with the
$Dp$ brane. As it moves in the throat, the probe emits closed strings
which carry away energy and the R-R charge.
It was found in \cite{sahakyan},\cite{nakayama1} that the  classical
analysis is only valid for $5> p >2$,
since the energy loss is not divergent in this instance. This will be assumed throughout.

Recent work \cite{kutasov2}, \cite{kluson2}, has emphasised that probing the compact
dimension of the $\mathbf{R^3 \times S^1}$ geometry is far better understood in terms
of tachyon condensation. However, we expect that the effective DBI action is more useful
for probing  $\mathbf R^3$. It is important to understand the motion of branes in
time dependent compactified backgrounds, since there may be implications for brane world cosmology.
\cite{mirage}, \cite{burgess}, \cite{yavartanoo}.

The purpose of this  work will be to attempt to address some of the
questions raised in \cite{kutasov}
regarding bound orbits and critical angular momenta.
We will begin with a review of the background solutions and the effective action for a BPS probe $Dp$-brane.
We will then briefly look at the behaviour of
the 'tachyon' potential before moving on to discuss the dynamics of the probe
brane, both with and without angular momentum, where we will try to explicitly solve the equations of motion.
We will also briefly investigate what happens when the compact dimension is not infinitesimally small,
and its effect on the probe brane. We conclude with some brief remarks and suggestions for future work.

\section{Effective action and background solution.}
We want to investigate the dynamics of a BPS $Dp$-brane in the background
of $k$ $NS5$-branes which are located at the origin of the transverse space-time
$\mathbf S^1 \times \mathbf R^3$. To this end, we must first remind ourselves of the
CHS solutions for the metric, dilaton and $NS$-3 form, which are given by \cite{chs},
\begin{eqnarray} \label{eq:chssolution}
ds^2&=& \eta_{\mu \nu} dx^{\mu} dx^{\nu} + H(x^m) \delta_{mn}dx^m dx^n
\nonumber\\
e^{2(\phi-\phi_0)}&=&H(x^m)\nonumber \\
H_{mnp}&=&-\epsilon^q_{mnp} \partial_q \phi
\end{eqnarray}
As usual $H_{mnp}$ is the field strength for the $NS$ B-field, whilst
$H(x^m)$ is the harmonic function describing the location of the $k$
fivebranes in the transverse space. Generically, the harmonic function
can be written as
\begin{equation}
H=1+ \sum^k_{i=1}\frac{l_s^2}{|\textbf x-\textbf x_i|^2},
\end{equation}
which, for coincident branes, becomes
\begin{equation}
H=1+\frac{k l_s^2}{r^2},
\end{equation}
where $r=\sqrt{x^m x_m}$ represents the radial distance from the source branes
in the transverse space $\mathbf R^4$.
We wish to consider a new background for the fivebranes, which we can create by compactifying
one of the transverse directions into a circle of radius $R_c$ (see \cite{peet},\cite{johnson} for example). We are free
to arbitrarily choose a direction to compactify since there exists an
$SO(4)$ symmetry in the transverse space. If we parameterise our chosen
direction by $y$ then the harmonic function can be written:
\begin{equation}\label{eq:harmonic_definition}
H=1+kl_s^2 \sum^\infty_{n=-\infty} \frac{1}{r^2+(y-2\pi R_cn)^2}
\end{equation}
If we now assume that the compactification radius is very small, then we
immediately see that this summation changes only very slowly with $n$ and
can be approximated by an integral
\begin{equation}
H=1+kl_s^2 \int \frac{dn}{r^2+(y-2 \pi R_cn)^2}.
\end{equation}
Performing a change of variables $ru=2 \pi R_cn - y$ and integrating over $u$
gives us the modified harmonic function,
\begin{equation}\label{eq:harmonic1}
H=1+\frac{\bar{k}}{r},
\end{equation}
where for simplicity we have defined $\bar{k}=kl_s^2/2R_c$.
This has immediate consequences for the validity of the solutions, since
we must now ensure that the following constraints are satisfied for the supergravity solution to hold.
\begin{equation}\label{eq:supergravity_constraints}
r>> R_c \hspace{1cm} kl_s^2 >> r R_c.
\end{equation}
If the radial distance $r$ approaches $R_c$ then this solution will no longer
be valid. We can still find an expression in this instance, since although
the denominator in the integral will no longer be a smoothly changing function, we can still
use the summation equation (\ref{eq:harmonic_definition}). This will yield a new harmonic function valid
deep in the CHS throat which will match onto the $1/r$ function at some
distance. We will refer to this 'deep throat' approximation later in the
paper. For now, we will be concerned with the geometry parameterised by
(\ref{eq:harmonic1}).

Into this background we wish to introduce a BPS Dp-brane, which is
extended in the $x^1 \ldots x^p$ directions and parallel
to the $NS$5 branes. As is usual in this situation,
we label the world-volume coordinates by $\zeta^{\mu}$ and go to static
gauge where $\zeta^{\mu}=x^\mu$. The embedding of the D-brane in the
transverse space gives rise to scalar fields on its world-volume, labelled
by $X^6(\zeta^{\mu}) \ldots X^9(\zeta^{\mu})$, and consequently we find
that the brane dynamics may be effectively described by the Born-Infeld
action:
\begin{equation}\label{eq:dbi}
S=-\tau_p \int d^{p+1} \zeta e^{-(\phi-\phi_0)} \sqrt{-det(G_{\mu\nu}+B_{\mu\nu}+2\pi l_s^2 F_{\mu\nu})}
\end{equation}
where $G_{\mu\nu}$ and $B_{\mu\nu}$ are the pullbacks of the metric and the
B-field to the branes world volume, whilst $F_{\mu\nu}$ is the $U(1)$ gauge
field and $l_s$ is the string length.
In this note we are interested in the case where the only excited field
on the brane is $r(\zeta^{\mu})$, and so the $B$-field couplings vanish.
Furthermore,
we will only be interested in homogeneous solutions where the scalar fields
are explicitly time dependent only. In this case the induced metric becomes
\begin{equation}
G_{\mu\nu}= \eta_{\mu\nu}+\delta_{\mu}^0 \delta_{\nu}^0 \dot{X}^m \dot{X}^m H(x^n).
\end{equation}
Inserting this into (\ref{eq:dbi}), and also including a non zero, constant
electric field on the brane gives the resultant D-brane action:
\begin{equation}\label{eq:action}
S=-\tau_p \int d^{p+1} \zeta \sqrt{H^{-1}(1-\lambda \varepsilon^2)- \dot{X}^m \dot{X}^m}
\end{equation}
where $\lambda$=$2 \pi l_s^2$, and $\varepsilon$ is the constant electric
field strength on the world-volume. Note that the DBI action can be used
here since it accurately describes the classical open string motion. It
will not however, account for the radiative corrections due to closed string
emission \cite{sahakyan}.
\section{Tachyon map.}
Tachyons in open string models are described by a Born Infeld action, which is given by \cite{garousi}
\begin{equation}
S=-\int d^{p+1} \zeta V(T)\sqrt{1+\partial_{\mu}T \partial_{\nu}T},
\end{equation}
where $V(T)$ is the tachyon potential. Upon comparison with (\ref{eq:action})
we find that setting the electric field to zero there is a map from one to the other, provided we define a
tachyon field
\begin{equation}
\frac{dT}{dr}=\sqrt{H(r)}=\sqrt{1+\frac{\bar{k}}{r}}.
\end{equation}
And consequently we know that the potential can be written as
\begin{equation}\label{eq:potential}
V(T)=\frac{\tau_p}{\sqrt{H(r)}}.
\end{equation}
With this definition we find
\begin{equation}
T(r)=\sqrt{r^2+\bar{k}r}+\frac{1}{2}\bar{k}ln(\frac{1}{2} \bar{k}+r+\sqrt{r^2+\bar{k}r}).
\end{equation}
Ideally we want to invert this equation to obtain an exact expression for the
tachyon potential, however this is not always possible and so we must
study the effective potential behaviour in its asymptotic limits. To leading
order, we find that
\begin{equation}
T(r\to R_c) \propto \frac{\bar{k}}{2}ln(\frac{\bar{k}}{2})
\end{equation}
\begin{equation}
T(r \to \infty) \propto r.
\end{equation}
We find that as $r\to R_c$, $T\to$ constant dependent upon the number of branes and the compactification radius. In the other
limit  we see that as
$r\to \infty$, $T \to \infty$ as expected.
Using this and also (\ref{eq:potential})
we find the effective behaviour of the tachyon potential.

\begin{equation}\label{eq:tachyon_potential}
\frac{V(T)}{\tau_p} \propto 1- \frac{\bar{k}}{2T} \hspace{0.5cm} T \to \infty.
\end{equation}

Thus as $T\to \infty$ we see that the potential changes as $1/T \propto 1/r$
as expected from considering the gravitational attraction. This agrees nicely with the solution
in the uncompactified case, where it was found that

\begin{equation}\label{eq:tachyon_potential_2}
\frac{V(T)}{\tau_p} \propto \Bigg \lbrace
\begin{array}{cc} e^{T/\sqrt{kl_s^2}} & T\to -\infty \\
  1- \frac{kl_s^2}{2T^2}&  T \to \infty
\end{array}
\end{equation}

Looking at the $T \to \infty$ case, we find that if $R_c=T$ then the solution in (\ref{eq:tachyon_potential}
is mapped to the solution in (\ref{eq:tachyon_potential_2}). This is because letting $T$ be large effectively 'blows up' the
compact dimension and we therefore recover our $SO(4)$ symmetry in the transverse space. Thus we should expect to recover the
Kutasov result \cite{kutasov}. The behaviour of the tachyon field as $r \to R_c$ is logarithmic, which suggests that the
potential would have some form of exponential behaviour (\cite{kutasov}, \cite{sen2}), albeit different from that in the
uncompactified case.
Geometrically we argue that the tachyon field is related to the radial
distance between the $D$-brane probe and the fivebranes, and therefore
by considering the brane dynamics we are learning about the behaviour of the
tachyon.
\section{Probe brane dynamics.}
We are now in a position to consider the dynamics of the probe brane in this
compactified background. Referring back to (\ref{eq:action}) we note that we
can use the $SO(3)$ symmetry to rotate the solution to the $x_8-x_9$ plane
by switching to polar coordinates. Namely, $x_8=Rcos\theta$ and $x_9=Rsin\theta$. The action now reads;
\begin{equation}
S=-\tau_p \int  d^{p+1} \zeta \sqrt{H^{-1}(1-\lambda \varepsilon^2)-\dot{R}^2-R^2\dot{\theta}^2}
\end{equation}
and we can construct the canonical momenta and canonical energy as usual. Note that now, $R$, is the radial distance in this plane and should not be confused with the compactification radius $R_c$.
\begin{equation}
\tilde{\Pi} = \frac{\Pi}{m}=\frac{\dot{R}}{\sqrt{H^{-1}(1-\lambda \varepsilon^2)-\dot{R}^2-R^2\dot{\theta}^2}}
\end{equation}
\begin{equation}
\tilde{L}=\frac{L}{m}=\frac{R^2\dot{\theta}}{\sqrt{H^{-1}(1-\lambda \varepsilon^2)-\dot{R}^2-R^2\dot{\theta}^2}}
\end{equation}
\begin{equation}
\tilde E =\frac{1}{H \sqrt{H^{-1}(1-\lambda \varepsilon^2)-\dot{R}^2-R^2\dot{\theta}^2}}.
\end{equation}
In deriving these expressions we have used the fact that $m=\tau_p \int d^p \zeta$ represents the effective 'mass' of the
brane.

In order to solve the equation of motion for $R$ in terms of fixed energy and 
angular momentum densities, we need to solve the angular momentum term
for $ \dot{\theta}$ and substitute the solution into the radial momentum
equation. The resultant expression is
\begin{equation}
\dot{R}^2=\frac{(1-\lambda \varepsilon^2)}{H}-\frac{1}{E^2 H^2}\left(1+\frac{\tilde{L^2}}{R^2}\right).
\end{equation}
Since this expression is non negative, it places constraints upon the 
strength of the electric field, which can be seen by simply substituting in
the expression for the harmonic function. Following recent work on
similar problems, it is convenient to define the effective potential to be
minus the radial equation of motion.

Using our expression for the probe brane action we can calculate the 
energy momentum tensor associated with it. Considering only the time 
dependent case we find the non zero components of the tensor are:
\begin{equation}
T_{00}=\frac{\tau_p}{H\sqrt{H^{-1}(1-\lambda\varepsilon^2)-\dot{R}^2-R^2\dot{\theta}^2}}
\end{equation}
\begin{equation}
T_{ij}=-\tau_p \delta_{ij}\sqrt{H^{-1}(1-\lambda \varepsilon^2)-\dot{R}^2-R^2\dot{\theta}^2}.
\end{equation}
As we can see, the time component of the tensor is associated with the
energy density on the brane. The spatial component can be rewritten in
terms of this density as;
\begin{equation}
T_{ij}=\frac{-\delta_{ij} \tau_p}{H\tilde E}.
\end{equation}
For small $R$, which we are calling the near horizon geometry, this can be written as
\begin{equation}
T_{ij}\propto \frac{-\delta_{ij} \tau_p R}{\bar{k}\tilde E}
\end{equation}
which tends to zero linearly with distance, in contrast to the original case
which vanished quadratically. It must be remembered that this is only valid
for $R >> R_c$, and so we cannot map this solution to the Kutasov one \cite{kutasov} in this
region.
\subsection{Radial motion in the near horizon limit.}
In this section we consider purely radial motion of the probe brane, and so
we drop the angular momentum terms. The constraint equation reduces to:
\begin{equation}
\frac{\bar{k}(1-\lambda \varepsilon^2)}{R}-\lambda \varepsilon^2 \ge \frac{1}{\tilde{E^2}}-1
\end{equation}
This tells us that $\lambda \varepsilon^2$ must be bounded above by unity, which is automatically satisfied since we have considered the field to be a small perturbation in order to 
derive the $D$-brane action.
The effective potential in this instance reduces to,
\begin{equation}
V_{\rm eff}=\frac{1}{\tilde E^2 H^2}-\frac{(1-\lambda \varepsilon^2)}{H}
\end{equation}
which is plotted in fig 1. It shows that the potential is attractive for small
energy density, becoming more repulsive as the energy increases.
We also see there are minima in the potential at large radial distances, which
could give rise to bound orbits. The effect of the increasing energy is to
move the minimum to larger distances from the fivebranes.

We can extract useful information about the 
behaviour of the probe by considering the limit of small and large $R$. 
As $R\to 0$, we neglect the factor of unity in the harmonic function since the
probe is located inside the throat of the geometry. As the probe travels down
the throat, is moves into a region of stronger coupling and so
we expect our perturbative solutions to break down. But we already know that this is the case, since at
very small $R$ the probe will 'feel' the effect of the compactification.
The potential in this region can be approximated by
\begin{equation}
V_{\rm eff} \propto \frac{R^2}{\tilde E^2 \bar{k}^2}-\frac{R(1-\lambda \varepsilon^2)}{\bar{k}}
\end{equation}
which can be seen to tend to zero as the radial distance decreases for arbitrary
energy density. We also note that the potential is identically zero
at the radial distance $R=\tilde E^2 \bar{k} (1-\lambda \varepsilon^2)$. This implies that there is some turning point in the potential for a certain range of values of energy and the electric field. It also implies that due to the 
supergravity constraints (\ref{eq:supergravity_constraints}) we must have $\tilde E^2 (1-\lambda \varepsilon^2) << 1$.

In this
region, we can solve the equation of motion explicitly. Upon integration we 
find that
\begin{equation}\label{eq:solr}
R(t)=\frac{(1+tan(y)^2 \pm \sqrt{tan(y)^2+tan(y)^4)}) \bar{k} \tilde E^2 (1-\lambda \varepsilon^2)}{2(1+tan(y)^2)}
\end{equation}
where we have defined $y=t/(\tilde E \bar{k})$.

We note that the solution above with - sign corresponds to 
an inwardly moving probe with initial (ie maximum ) 
distance from the $NS$5-branes, given by $R= \bar{k} \tilde E^2 (1-\lambda 
\varepsilon^2) / 2$.  Temporarily setting
the electric field to zero we see this implies that the smaller we
make $R_c$ then the further away the probe will be. By turning
on the electric field we see that this has an affect on the maximum
distance. In fact we note that this distance decreases as
we increase the strength of the electric field. Thus for a fixed value of $R_c$ we see that the increasing field strength puts
the probe brane
closer to the fivebranes.

Another point worth noting is that since tan functions are defined
such that the argument lies between $-\pi/2$ and $\pi/2$, 
for a fixed energy density, there are strict constraints on the time 
evolution of our solution. This is to be expected since we know that the 
probe must be deep in the fivebrane throat. What the solution tells us 
though, 
is that as $y \to \pi/2$, $R(t) \to 0$. This means that the probe
 will reach the source branes in a finite time as measured from the 
fivebranes. This is interesting since it contrasts with
the uncompactified case, where that particular solution suggested that
 it would take an infinite amount of time as measured by the fivebrane, 
but only a finite time when measured using the proper time on the $D$-brane.

Even though the minus sign solution corresponds to probe motion toward the 
fivebranes we know that in the $R \to 0$ limit the theory will break down 
since the compact dimension will
in essence `decompactify'. Furthermore, there will be strong stringy 
effects to
take into account, as discussed by \cite{kutasov}, \cite{sahakyan}.

For the solution with positive sign in (\ref{eq:solr} ) , corresponding 
to 
motion away from 
the
fivebranes, we see that the probe potentially starts at some 
\emph{minimum} distance. With time evolution it moves away to some maximum 
distance before returning
to its original position. However at this point, it would be expected to
match onto the in-falling solution. The maximum distance for the outgoing probe is given by
\begin{displaymath}
R_{\rm max}= \bar{k}\tilde E^2 (1-\lambda \varepsilon^2),
\end{displaymath}
which is also the place where the effective potential vanishes. The difficulty with this solution is that
it is unphysical. A probe brane (with electric field) moving into the throat from Minkowski space will either have too much energy
and simply head toward the fivebranes, or will oscillate around the minimum of the gravitational potential.

In the large $R$ approximation we know that the background spacetime is
flat Minkowski, and so the effective potential goes as
\begin{equation}
V_{\rm eff} \propto \frac{1}{\tilde E^2}-(1-\lambda \varepsilon^2)
\end{equation}
which is large and positive for $\tilde E << 1$, and small and negative
for $\tilde E >> 1$. Thus we see that the potential is attractive for
small energy density and repulsive for large energy density. Combining
this with what we know from the throat approximation, we note
that for $\tilde E <1$ the potential will lead to a bound state since it
has a crossing point at a non zero value of $R$. In terms of the tachyon
description we see that the $D$-brane will be unstable and will roll down the
potential. For $\tilde E > 1$ the potential is repulsive and will not lead to the formation of a bound state, which
is to be expected since we know that this will violate the supergravity constraints.
The effect of an increased electric field strength is to shift the potential, reducing the depth of the minimum and
bringing it closer to the fivebranes. This is shown in figure 2.

The explicit solution of the equation of motion in this region is given by
\begin{equation}
R(t)=t\sqrt{(1-\lambda \varepsilon^2)-1/\tilde E^2},
\end{equation}
which shows that the probe brane approaches the throat linearly with
time, as measured by an observer on the fivebranes.
\subsection{Combined radial and angular motion.}
With the inclusion of the angular momentum density, we expect to see slightly altered dynamics.
We again investigate the potential in the two limits as in the previous 
section. Firstly, for motion in the throat the effective potential is;
\begin{equation}\label{eq:angular1}
V_{\rm eff} \propto \frac{R^2}{\tilde E^2 \bar{k}^2}\left(1+\frac{\tilde{L}^2}{R^2}\right)-\frac{R(1-\lambda \varepsilon^2)}{\bar{k}}
\end{equation}
which does not vanish as $R\to0$, in fact it tends to a constant dependent upon
the energy density and the angular momentum density. In all cases we find that
the potential is repulsive in the throat, independent of the size of the
energy density.
(\ref{eq:angular1}) also provides us with another constraint on the allowed values of the energy
density and angular momentum. It turns out that they must satisfy 
\begin{equation}
\frac{1}{\tilde E^2 (1-\lambda \varepsilon^2)}\left(1+\frac{\tilde L^2}{R^2}\right) >> 1.
\end{equation}
The solution to the equation of motion in the throat is given by
\begin{equation}\label{eq:eom2}
R(t)=\frac{\tilde E^2 \bar{k}(1-\lambda \varepsilon^2)(1+tan(y)^2)\pm \sqrt{\tilde E^4 \bar{k}^2(1-\lambda \varepsilon^2)B-4\tilde{L}^2B}}{2(1+tan(y)^2)}.
\end{equation}
Where $y$ is defined as before, and $B$=$tan(y)^2+tan(y)^4$. Once again we
see that at $t$=0, the probe is located at $\bar{k} \tilde E^2 (1-\lambda \varepsilon^2)/2$. Choosing the minus sign again corresponds to inward motion. We see that the solution describes some kind of oscillatory motion, but now there is 
the additional angular momentum term to consider.

There appears to be a critical value for the angular momentum emerging from this solution, and the vanishing of the potential.
\begin{displaymath}
\tilde L \le \frac{\tilde E^2 \bar{k} \sqrt{1-\lambda \varepsilon^2}}{2}.
\end{displaymath}
If the angular momentum satisfies this condition, then the solution is valid.
 However in the limit that the LHS is equal to the right hand side,
then the square root term in (\ref{eq:eom2}) vanishes. This means that the
 whole equation for $R$ is
independent of time, and so the probe brane will be at a constant distance
 from the fivebranes. Physically we can think of it as rotating rapidly around
 the throat. The position of the
probe in this case is dependent upon the size of $R_c$ and the electric field
 strength. This suggests that if $\tilde E < 1$ then the angular momentum term
 is vanishingly small, and can really only be defined for the $\tilde E >1$
 case. If the constraint is violated then we pick up an imaginary term in 
the equation of motion.
If we look at the vanishing of the potential, we find the solutions are
\begin{equation}
R_0 = \frac{\tilde E}{2} \left( \tilde E(1-\lambda \varepsilon^2) \pm \sqrt{(1-\lambda \varepsilon^2)^2 \tilde E^2 - 4 \tilde L^2} \right).
\end{equation}
Thus when there is no angular momentum we see that the zero of the potential occurs at $R_0 = 0$ and/or $R_0 = \tilde E^2(1-\lambda \varepsilon^2)$ as
expected. For non-zero angular momentum we again find our constraint condition. Provided the constraint condition is not broken, we see that there
will be zeros of the potential and thus bound orbits will exist.

Now we are in a position to discuss the probe motion in the throat. We have
already seen that for $\tilde L = 0$, the probe either passes through
the fivebranes or is bound to some maximum distance. This is modified 
somewhat with the addition of the angular momentum term. We now find that;
\begin{equation}
R(t\to \pi/2) = \frac{\tilde E^2 \bar{k}(1-\lambda \varepsilon^2) \pm \sqrt{\tilde E^4 \bar{k}^2 (1-\lambda \varepsilon^2) - 4\tilde L^2}}{2}
\end{equation}
where the angular momentum constraint must be satisfied for the solution to remain real. In this instance, for an outgoing probe, we find the same behaviour as
we did for purely radial motion. The probe starts off at some minimum
distance and moves outward to a maximum distance which is dependent on the strength of the energy density and the electric field, before returning
to the minimum. If we now consider the in-falling brane, we find different
behaviour. The probe starts off at some maximum distance and travels down the throat toward the fivebranes. If the energy is sufficiently large
the probe will reach $R=0$. If not, then it will be in a bound state with the fivebranes. The exact position of the bound state in the throat
will be dependent upon the strength of the electric field.

In the large $R$ limit we see that the effective potential reproduces the
purely radial one, since the angular momentum becomes negligible at large 
distances (in fact the angular momentum is negligible for
all distances much larger than the compactification radius as can be seen in figure 3.)

In this case the equation of motion can again be solved explicitly with the
solution
\begin{equation}
R(t)= \frac{\sqrt{t^2 \tilde E^4 (1-\lambda \varepsilon^2)^2-2\tilde E^2 
t^2
(1-\lambda \varepsilon^2)+t^2+\tilde E^2 \tilde L^2}}{( 
\tilde{E} \sqrt{(\tilde{E}^2(1-\lambda \varepsilon^2)-1)} )}
\end{equation}

The effect of the electric field in this case will
be to alter the position of the turning point, moving it toward zero as the
field strength increases.

Using our canonical momenta we can find the equation of motion for the
angle $\theta$ in terms of a fixed energy density
\begin{equation}
\dot{\theta}=\frac{\tilde L}{\tilde E H R^2}.
\end{equation}
Far from the throat, where $H\to1$, we find that
\begin{equation}
\dot{\theta}= \frac{\tilde L}{\tilde E R^2}
\end{equation}
and upon integration we have
\begin{equation}
\theta= \frac{\tilde E^2 arctan(Qt/\tilde E \tilde L)-2\tilde E^4 arctan(Qt/\tilde E \tilde L)+\tilde E^6 arctan(Qt/\tilde E \tilde L)}{Q}.
\end{equation}
Where we have defined $Q=\sqrt{\tilde E^4 (1-\lambda \varepsilon^2)^2-2\tilde E^2(1-\lambda \varepsilon^2)+1}$. This shows that the angle initially
changes rapidly but then tends toward a fixed value as time evolves.
However in the throat we can use our standard approximation of the harmonic
function and we find
\begin{equation}
\dot{\theta}=\frac{\tilde L}{\tilde E \bar{k} R}.
\end{equation}
This can be integrated, but the resulting expression is extremely complicated.
We perform a Taylor expansion for small $t$ to give us a description of the early time behaviour, which shows that to leading order
the angle changes logarithmically. This is different to the uncompactified
case \cite{kutasov}, where it was found that the probe spiralled around the throat toward the fivebranes.

If these bound orbits can occur, we want to know what their trajectories
are. Fortunately we are able to make some headway with this since the
orbit of the probe brane in the near horizon geometry is given by:
\begin{equation}
\phi-\phi_0 = \int \frac{dx}{\sqrt{-x^2 + \tilde{B}x + C}}
\end{equation}
Where we have defined $x=1/R$ as is customary, and also
\begin{equation}
\tilde{B} = \frac{\tilde{E^2} \bar {k} (1-\lambda \varepsilon^2 )}{\tilde 
L^2}
\nonumber 
\end{equation}

\begin{equation}
C = \frac{-1}{\tilde L^2}.
\end{equation}

This allows us to solve the integral, which up to arbitrary constants, becomes
\begin{equation}
tan(\phi) = \frac{x-1/2 \tilde{B}}{\sqrt{-x^2+ \tilde{B}x+C}}.
\end{equation}
Upon expansion this yields a simple quadratic equation and so we obtain the explicit solution for the radius of the orbit,
\begin{equation}
R = 2 \left( \tilde{B} \pm sin(\phi) \sqrt{\tilde{B}^2+4C}\right)^{-1}.
\end{equation}
The first thing to note is that there will be an orbit  of constant radius, ie circular, if
\begin{equation}
\tilde L =\frac{\tilde E^2 \bar{k}(1-\lambda \varepsilon^2)}{2},
\end{equation}
which is the solution we obtained from the equations of motion. Provided that
the angular momentum constraints are fulfilled, we appear to have elliptic orbits parameterised by
\begin{equation}
a = \frac{\tilde L^2 \tilde E^2 \bar{k} (1-\lambda \varepsilon^2)}{2}, \hspace{0.5cm} e = \sqrt{1-\frac{4\tilde L^2}{\tilde E^4 \bar{k}^2 (1-\lambda \varepsilon^2)^2}}.
\end{equation}
where, as usual, $a$ is the semi-latus rectum, and $e$ is the eccentricity of the orbit. These orbits are just the standard conic sections one would expect in
this background.
\subsection{Deep throat region.}
We will now examine what happens as the probe brane reaches a  distance that
is comparable with the compactification radius. In this instance we must 
resort to (\ref{eq:harmonic_definition})
\begin{displaymath}
H=1+kl_s^2 \sum^\infty_{n=-\infty} \frac{1}{Z^2+(Y-2\pi R_cn)^2},
\end{displaymath}
where we have chosen to parameterise the coordinates on the $S^1$ by $Y$, and 
the transverse coordinates in $\mathbf R^3$ by $Z$ for convenience.

Since this solution will be valid deep in the CHS throat, we choose to rescale
the distances by a factor of $g_s$, and then send $g_s \to 0$ whilst keeping
the rescaled distances fixed - which is the limit taken when discussing
Little String Theory \cite{aharony}.
\begin{equation}
Y=g_sy; \hspace{0.5cm} Z=g_sz; \hspace{0.5cm} R_c= g_s r_c
\end{equation}
The harmonic function in terms of these rescaled distances is given by
\begin{equation}\label{eq:rescaled_harmonic}
H(z,y)= \frac{kl_s^2}{2r_c |z|} \frac{ sinh(|z|/r_C)}{( 
cosh(|z|/r_c)- cos (y/r_c))}.
\end{equation}
where $y$ ranges from $0 \ldots \pi r_c$. This parameterisation is chosen
to ensure that the fivebranes sit at $y=0$.
At relatively large transverse distance, we can neglect the $y$ term which implies that
the harmonic function becomes
\begin{equation}
H(z,0) \approx \frac{kl_s^2}{2|z|r_c} tanh(|z|/2r_c),
\end{equation}
whilst at very small distances (or equivalently when the z field is minimized) we find
\begin{equation}
H(0,y) \approx \frac{kl_s^2}{4r_c^2  sin(y/2r_c)^2}
\end{equation}
The behaviour of a probe brane with this harmonic function was examined
in \cite{kutasov2}, where the relationship between BPS and Non-BPS branes
was elucidated. The probe brane appears to be stable if placed at $y=\pi r_c$, but this is in fact a point of instability. As a result, the probe will be attracted toward the fivebranes located at $y=0$ or $y=2\pi r_c$. This can be viewed in terms of the rolling tachyon using the tachyon map described in an earlier
section.
Note that if we send $y \to 0$ we recover the usual $SO(4)$ symmetry associated
with the uncompactified case.

The action for the probe brane in this background is similar to the one in the
earlier sections, except that we have a rescaled tension $T_p = g_s \tau_p$, 
and we can consider velocity in the $z$ and $y$ directions.
The rescaled energy in this region is given by
\begin{equation}
\tilde E = \frac{1}{H(z,y)\sqrt{H^{-1}(z,y)(1-\lambda\varepsilon^2)-(\dot{z}^2+\dot{y}^2)}},
\end{equation}
and the equations of motion for fixed energy are given by
\begin{equation}
\dot{z}^2+\dot{y}^2 = \frac{(1-\lambda\varepsilon^2)}{H(z,y)}-\frac{1}{\tilde E^2 H^2(z,y)}.
\end{equation}
We know that for distances much larger than $r_c$ we can use the results from the previous section, whilst for distances smaller than $r_c$ we effectively
have a dimensionality crossover where the harmonic function behaves as $1/r^2$. Thus, it is useful to
investigate what happens when the probe brane is at a distance comparable
with the compactification radius.

From the equations of motion, we have the non-negativity constraint
\begin{equation}
\frac{kl_s^2(1-\lambda\varepsilon^2)}{2|z|r_c}\frac{ sinh(|z|/r_c)}{(
 cosh(|z|/r_c)- cos(y/r_c))} \ge \frac{1}{\tilde E^2}.
\end{equation}
If we assume that $z \approx r_c$ we can perform a Taylor series expansion which yields:
\begin{equation}
\frac{kl_s^2(1-\lambda\varepsilon^2)}{2z^2}\frac{1/2(e-e^{-1})}{
1/2(e+e^{-1})- cos(y/r_c)} \ge \frac{1}{\tilde E^2}.
\end{equation}
Since the cosine term is constrained to be between $\pm 1$, we find that the 
constraint can be written
\begin{equation}
\frac{kl_s^2 \mathcal{O}(1)}{2z^2}\ge \frac{1}{\tilde E^2}.
\end{equation}
As before we write the effective potential as
\begin{equation}
V_{\rm eff} = \frac{1}{\tilde E^2 H^2(z,y)}-\frac{(1-\lambda\varepsilon^2)}{H(z,y)}
\end{equation}
which we plot in figures 5 and 6. As we can see, in the region of validity we
find that the potential is repulsive for small values of $\tilde E$. The 
minimum potential in this region is at the point $y=\pi r_c, z=0$ as expected
from \cite{kutasov2}. 
The effect of the electric field in this instance is to reduce the height of the potential 
along the $\mathbf S^1$ direction.

\section{Discussion.}
We have seen that by compactifying one of the transverse dimensions to the 
$NS$5-brane, we do in fact find bound orbits at sufficiently large distances.
The inclusion of an electric field on the probe brane also has a strong
effect on the dynamics, since we have found that  an increasing 
electric field strength tends to localise the probe brane nearer the fivebranes - i.e in a region of strong coupling.
 This is what we would expect, since the field 
is effectively adding more 'mass' to the brane and so we would expect it to sink
further into the throat.

Using the tachyon map we have explored the differences between the solutions in a compactified background and the general background, and also seen how they
can be mapped to one another. Interestingly we find that there is a unique 
size of the compactification radius, namely $R_c = \sqrt{k l_s^2}/2$, which
will map the compactified solution to the uncompactified one as $T\to -\infty$.
It would be interesting to understand the underlying reasons for this.

When we include the angular momentum term we expect to see different dynamics.
 However there are constraints emerging from the throat solution which 
suggest that the angular momentum term is negligible in the near horizon region.
If the angular momentum saturates the equation of motion constraint, then we will find that 
the probe executes circular motion in the throat at a distance determined by the compactification
radius and the electric field strength. Since the $D$-brane will constantly
be radiating energy as it moves, this trajectory is unlikely to be 
stable - however since the probe carries Ramond-Ramond charge there may
be something akin to synchrotron radiation emission \cite{savvidy}. 
The dynamics of the probe as it approaches the compactification 
radius have also been addressed.

In closing it appears that we have verified Kutasov's conjecture and also
seen the effect that an electric field has on the probe dynamics. There is still
the issue of the critical angular momentum to consider, since altering
the background geometry appears to have no effect on this constraint.
Further work must be done to determine the underlying reasons for this.
The work could also be extended to consider $Dp$-brane backgrounds and also
other types of intersecting brane backgrounds with relative ease.
Also the presence of a bound state could be useful for cosmology, in particular an
extension of the work already done in \cite{mirage}, \cite{yavartanoo} and \cite{Ghodsi}.
In relation to the work done in \cite{kutasov2}, it would also be of interest
to consider Non-BPS branes in this background.
Recent work on a related problem has appeared in \cite{kluson}, \cite{kluson3}.

\textbf{Added Note:} After completing this work we became aware of
ref \cite{chen2} which has a sizable overlap with the present paper.
However we have in addition, also presented in our work, analytic
solutions regarding the probe brane orbits.


%

\newpage
\begin{figure}\begin{center}
\includegraphics[width=10cm,height=7.4cm]{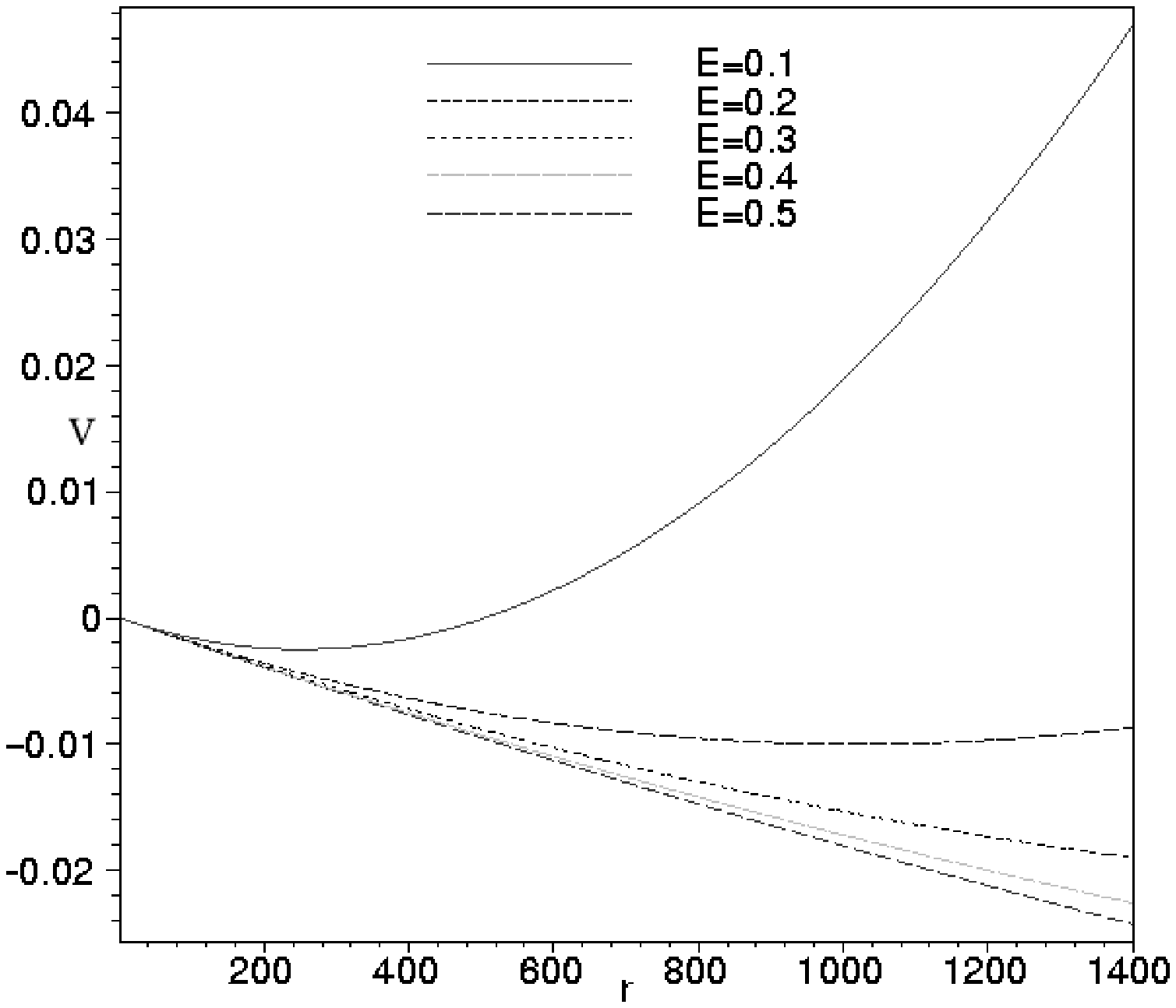}
\end{center}\caption{Potential for various energy density values with zero field strength. We have taken $k$=100000,$ l_s$=1 and $R_c$=1 for
simplicity.}
\end{figure}
\begin{figure}\begin{center}
\includegraphics[width=10cm,height=7.4cm]{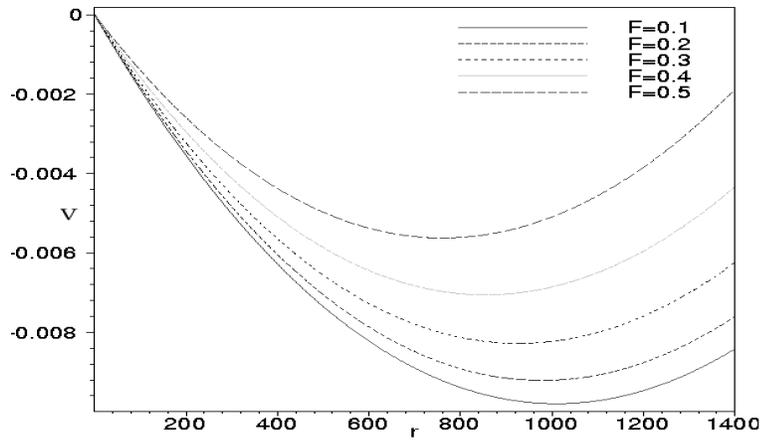}
\end{center}\caption{Potential for fixed $\tilde E<<1$
but with increasing (dimensionless) electric field strength. Note that the stronger
electric field shifts the minimum toward $r=0$.}
\end{figure}

\begin{figure}\begin{center}
\includegraphics[width=10cm,height=7.4cm]{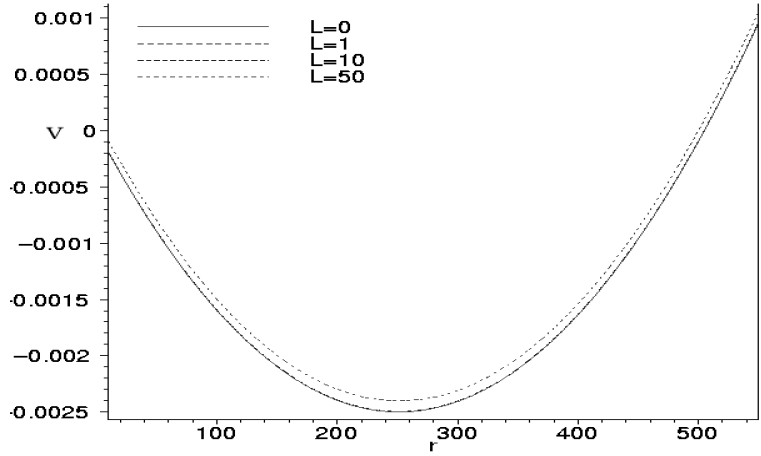}
\end{center}\caption{Potential for fixed $\tilde E<<1$, and zero electric flux. The
angular momentum appears to be negligible in this region.}
\end{figure}
\begin{figure}\begin{center}
\includegraphics[width=10cm,height=7.4cm]{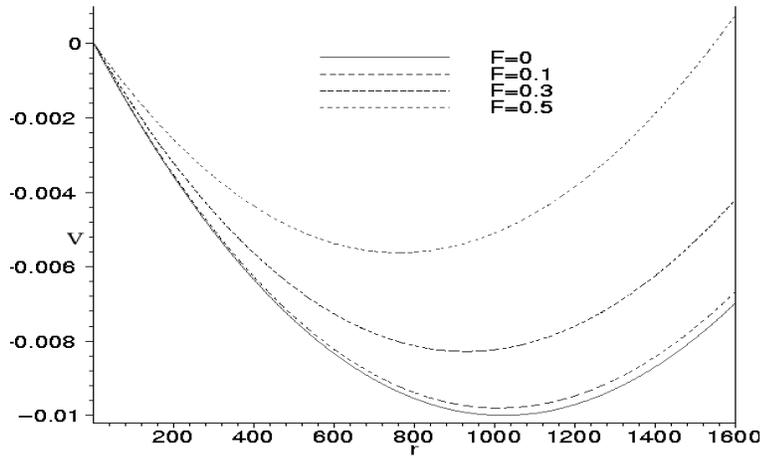}
\end{center}\caption{Potential with $\tilde L=1$ and fixed energy
density, $\tilde E <<1$. We see that the effect of the electric field is to
shift the position of the minima to smaller $r$.}
\end{figure}
\begin{figure}\begin{center}
\includegraphics[width=10cm,height=7.4cm]{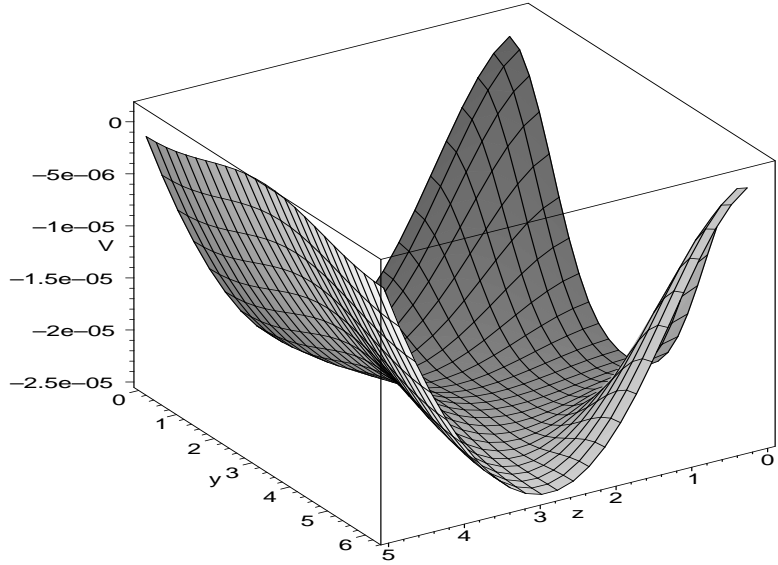}
\end{center}\caption{Potential for $\tilde E = 0.01$ as the probe moves into a region where it is comparable with the compactification radius. We have set the electric field to zero here.}
\end{figure}
\begin{figure}\begin{center}
\includegraphics[width=10cm,height=7.4cm]{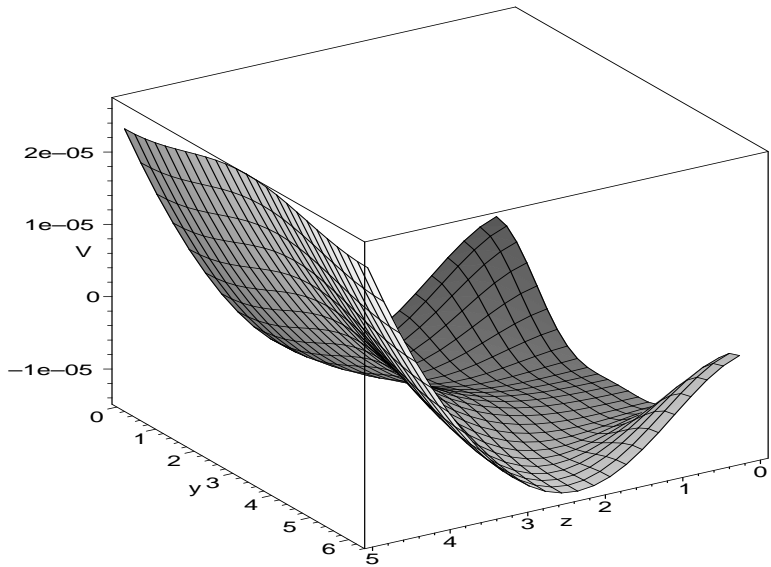}
\end{center}\caption{Potential for $\tilde E$=0.01 with an electric field strength of
F=0.5 showing how the electric field alters the potential along the circle when z=0.}
\end{figure}

\end{document}